# Reconstructed spatial receptive field structures by reverse correlation technique explains the visual feature selectivity of units in deep convolutional neural networks

Yoshiyuki R. Shiraishi [1,] Hiromichi Sato [1, 2,] Takahisa M. Sanada [3], Tomoyuki Naito [2*]
Laboratory of Cognitive and Behavioral Neuroscience, Graduate School of Frontier Biosciences, Osaka University
[1]Laboratory of Cognitive and Behavioral Neuroscience, Graduate School of Medicine, Osaka University, Japan
[2]Faculty of Software and Information Science, Iwate Prefectural University, Japan
[3]Laboratory of Cognitive and Behavioral Neuroscience, Graduate School of Medicine, Osaka University, Japan
*Corresponding author E-mail: naito@vision.hss.osaka-u.ac.jp

**Abstract**

An important issue in dealing with Deep Convolutional Neural Networks (DCNN) is the "black box problem", which represents the unknowns about internal information representation and processing, especially in the middle and higher layers. In this study, we adopted a systems neuroscience methodology to measure the visual feature selectivity and visualize the spatial receptive field of the units in VGG16. Orientation and spatial frequency tunings of each unit were measured using sinusoidal grating stimuli. The image category selectivity of each unit was also measured using natural image stimuli. The spatial structures of the receptive fields of all convolutional units were estimated by activation-weighted average (AWA) and activation-weighted covariance (AWC) analyses. In the middle layers (convolutional layers in block3 and block4), AWC analysis successfully reconstructed the receptive field that predicted the visual feature selectivity of the unit. Those results suggested the possibility that analyzing the reconstructed receptive field structure can be used to interpret the functional significance of the units and layers of a DCNN.

## 1 Introduction
### 1.1 Blackbox problem in artificial and biological neural networks

Deep convolutional neural networks (DCNNs) are classifiers that include a feature extractor and have recently been gaining prominence in the field of computational image recognition (Krizhevsky et al., 2012; Simonyan & Zisserman, 2014; Szegedy et al., 2015). The precursor of convolutional neural networks (CNNs), Neocognitron (Fukushima & Miyake, 1982), was elaborated to have a high capability of handwritten character recognition with backpropagation, as demonstrated by ConvNet (LeCun et al., 1989). After the success of AlexNet (Krizhevsky et al., 2012) at the 2012 ImageNet Large Scale Visual Recognition Competition (ILSVRC), CNNs became the preferred approach in image categorization or image detection competitions. CNNs have since evolved into DCNNs by the introduction of multi-stratified convolutional layers. The image classification ability of DCNNs is close to or beyond that of humans; for example, ResNet (He et al., 2016) exhibited a classification error rate of 3.57% (the human classification error rate is 5.1%) at ImageNet (Russakovsky et al., 2015).

When trying to apply DCNNs to tackle practical problems such as medical diagnostic imaging or detection tasks of specific persons, however, a "black box problem" is encountered (Chakraborty et al., 2017; Fan et al., 2020; Zhang & Zhu, 2017). That is, while the image classification or image detection accuracy is equal to or better than that of humans, how this ability is acquired and represented in the networks is unclear. In other words, there is no theoretical understanding of how and why DCNNs work even though the complete connectivity and activation patterns of DCNNs are fully accessible. The reason is sometimes attributed to the complexity of DCNNs. For example, AlexNet has a relatively shallow DCNN compared with modern DCNNs, which can typically have eight layers and a total of 60 million parameters (Krizhevsky et al., 2012).

Neuroscience faces a similar challenge. Although the biological neural networks of primates including humans can classify complex images rapidly, how visual neurons compute and represent visual information for image classification tasks remains controversial. This controversy may be attributed to the complexity of the brain, in which there are even more units and more connection complexity than in DCNNs: the human cerebral cortex has roughly about 16 billion neurons (Azevedo et al., 2009) and about 160 trillion synapses (Y. Tang et al., 2001).

For biological neural networks, a canonical analysis that provides insight into neural representations and computations has been used to reconstruct the receptive field structure of a single neuron. The spatial receptive field of a neuron is defined as the region of the stimulus space causing a neural response, which is usually characterized by the preferred stimulus that causes the maximum response of the neuron (Hartline, 1940; Hubel & Wiesel, 1959, 1968). In DCNNs, the so-called activation maximization, which corresponds to receptive field analysis in neuroscience, was used to analyze the units in neural networks (Erhan et al., 2009; A Nguyen et al., 2017; Anh Nguyen et al., 2016; Zhou et al., 2015). Activation maximization analysis revealed that the receptive fields of units in early layers frequently have center-surround or Gabor function-like spatial structures, which are very similar with the receptive field structures of neurons in the lateral geniculate nucleus (LGN) and primary visual cortex (V1), respectively. The units in intermediate layers often respond to complex textures, and those in higher layers respond to object-like images, which is consistent with the selectivity of neurons in intermediate and higher visual areas, respectively (Erhan et al., 2009). Although, activation maximization analysis allows us to evaluate the functional significance of the units and layers of DCNNs qualitatively, quantitative evaluation has remained difficult, especially at the middle and higher layers. For example, it is difficult to predict how strongly a unit responds to unknown stimuli by the activation maximization method.

### 1.2 Reverse correlation method

In this study, to evaluate the functional significance of DCNN units quantitatively, we adopted reverse correlation analysis. In this approach, neuronal responses to broadband stimuli (1D or 2D Gaussian white noise stimuli were frequently used) were collected. Then, by calculating the spike-triggered average (STA) of the stimuli, we could visualize the spatiotemporal receptive field structure of each unit as a linear filter (Meister et al., 1994; Reid & Alonso, 1995; Sakai & Naka, 1987). The STA has been used to reconstruct the receptive field of auditory neurons (Eggermont et al., 1983), retinal ganglion cells (Chichilnisky, 2001; Meister et al., 1994; Sakai & Naka, 1987), relay cells in the LGN (Reid & Alonso, 1995) and V1 simple cells (DeAngelis et al., 1993; Jones & Palmer, 1987). The STA can also predict the responses to grating stimuli or natural image stimuli well (Bonin et al., 2005; Suematsu et al., 2012, 2013). Additionally, it has been applied to membrane potentials instead of neural spikes (Sakai, 1992). In this study, we refer to an STA as an "activation-weighted average (AWA)", because DCNNs do not exhibit spiking responses.

### 1.3 Activation-weighted covariance analysis

A theoretical study demonstrated that the STA (and thus AWA) does not work for the analysis of complex cells in V1 or neurons in the higher visual area (Simoncelli et al., 2004). Complex cells represent information not only in terms of the mean response but also in terms of the variance of the responses (e.g. the neuron can respond more to brighter and darker stimuli than the average). As a result, the STA (and AWA) is canceled out, resulting in no significant spatial structures being reconstructed. To overcome this issue, a more elaborated method, spike-triggered covariance (STC) analysis, was introduced to visualize the receptive field structure of complex cells as an aggregation of excitatory and inhibitory linear sub-filters (Horwitz et al., 2005; Pillow & Simoncelli, 2006; O Schwartz et al., 2002; Odelia Schwartz et al., 2006; Touryan et al., 2002). The sub-filters provide information on the neuron's presynaptic functional connectivity and on how the neuron integrates sensory stimuli. Thus, the sub-filters of a neuron allow us to elucidate the functional significance of complex cells and/or neurons in the higher visual areas. In this study, we call STC analysis "activation-weighted covariance (AWC)" analysis, because, as explained above, DCNNs do not exhibit spiking responses.

Neurons in higher visual areas (e.g. V4 and inferior temporal cortex (IT cortex) in the ventral visual pathway) are often not responsive to white noise stimuli (Aljadeff et al., 2016). Furthermore, compared with neurons in lower visual areas, the response rate of neurons in higher visual areas is low, and the stimulus selectivity is complex. In general, to reconstruct linear sub-filters of the receptive field of a neuron with more complex stimulus selectivity requires more responses elicited by the stimuli. However, the collection of enough spiking responses for STC (and AWC) analysis in the higher visual cortex of a biological neural network is not trivial.

In contrast, with regards to DCNNs, we can collect an unlimited number of responses elicited by stimuli, which should therefore allow for AWC analysis even of the higher layers.

### 1.4 The aim of this study

In this study, we tackled the black box problem of DCNNs using AWA and AWC analyses to reconstruct the spatial receptive field structure of the convolutional units of a DCNN. We measured the orientation selectivity and spatial frequency selectivity of the convolutional units, because they are the most characteristic properties of visual neurons throughout the ventral stream of the visual cortex, especially the early visual cortex (Ferster & Miller, 2000; Hubel & Wiesel, 1959, 1968). We also measured the natural image category selectivity of each convolutional unit. It is known that visual neurons in the higher ventral visual area (i.e. IT cortex) exhibit significant image category selectivity, while early visual neurons do not show natural image selectivity (Tanaka, 1996). It was also reported that the convolutional units in the highest layer of a DCNN also exhibit image category selectivity (Richards et al., 2019; Yamins et al., 2014; Zhou et al., 2015). The specific steps to reconstruct the spatial receptive field and verification of validity are as follows:

1) The orientation and spatial frequency selectivity of the convolutional units were measured using grating stimuli.
2) The image category selectivity of each convolutional integrated unit was measured using natural image stimuli.
3) The spatial structures of the DCNN receptive field were visualized by AWA and AWC analyses.
4) The prediction accuracy of the reconstructed receptive fields was assessed using a linear-nonlinear cascade model (LN model).

The successful reconstruction of the spatial sub-filters of the receptive fields of units in intermediate and higher layers achieved in this study will help elucidate the functional significance of the units.

## 2 Method

### 2.1 Neural network architecture

In this study, the well-trained VGG16 (Simonyan & Zisserman, 2014) built on the popular machine learning library Keras (Chollet & others, 2015) was used as a simple DCNN that achieved an image recognition score equal to or higher than that of humans at ILSVRC 2014.

To calculate the visual feature selectivity, we defined the response magnitude of a unit as the center value of the feature map of the kernel (if a feature map size was 224× 224 pixels, the response magnitude was defined as the value of the feature map at [112, 112]). We recorded all unit responses of all 13 convolutional layers.

### 2.2 Orientation and spatial frequency tuning of DCNN units

We measured the orientation and spatial frequency tuning selectivity of a unit using Cartesian grating stimuli (four spatial phases: 0, 90, 180, and 270°; six spatial frequencies: equal intervals in logogrammatic scale ranging between 1.75 -56.0 cycles/image; and 17 orientations: 0 -170° in 10 degrees steps). The unit response $R(\theta_i, \omega_j)$ at orientation $\theta_i$ and spatial frequency $\omega_j$ was defined as a rectified average of the response magnitude of the four spatial phases. $R(\theta_i, \omega_j)$ was plotted in polar coordinates.

### 2.3 Natural image category selectivity of DCNN units

Natural images were obtained from a public image database: computational vision Caltech 101(Fei-Fei et al., 2006). We selected 40 images in five categories: 'bonsai', 'car', 'cup', 'face' and 'rooster'. These categories cover artifacts, natural objects and human faces and are often used as visual stimuli in visual neuroscience research. To verify a significant natural image selectivity, we used a one-way

ANOVA with Bonferroni correction (significance level, $p < 0.01$).

## 2.4 Reconstructing the DCNN spatial receptive field

Using the reverse correlation technique, we reconstructed the spatial receptive field structure of the DCNN units. Because feedforward DCNNs have no temporal properties of output responses, the receptive field of DCNN units only have a 2D spatial profile. We presented one million white noise stimuli to VGG16 and corrected the responses of each unit. The two different reverse correlation methods (AWA and AWC) were applied to each unit.

## 2.5 AWA analysis

To reconstruct the spatial profiles of the receptive field of the convolutional units, activation-weighted white noise stimuli were averaged as AWA (Fig. 1). In the case that the stimulus is spherically symmetric, the AWA is equivalent to estimating an unbiased receptive field From the point of view of system identification, the AWA is equivalent to the first order Wiener kernel or Volterra kernel under the assumption of a linear system. This concept was previously described as linear filters following a static nonlinear operation and Poisson spike generator (LNP) model (Chichilnisky, 2001).

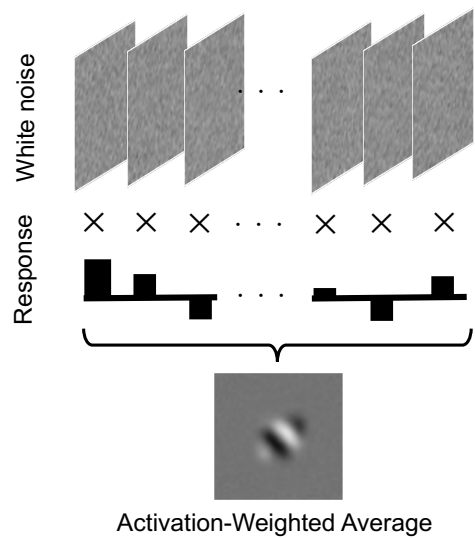

Figure 1. **Illustration explaining calculation of the activation-weighted average (AWA). The AWA of 1 million Gaussian white noise stimuli was calculated as an AWA filter.**

## 2.6 AWC analysis

However, there might be some units with multiple kernels such as V1 complex cells. For complex cells, LNP model expansion is required to obtain multiple linear kernels and their nonlinear combinations. One method for this purpose is STC (or AWC in this study) analysis. The AWC matrix often has more than two kernels that are provided as eigenvectors. The AWC kernels are equivalent to the second order Wiener kernel, and an eigenvalue indicates the weight of the kernel for the unit responses. An eigenvector with eigenvalues larger than the average corresponds to an excitatory kernel, and an eigenvector with eigenvalues smaller than the average corresponds to a suppressive kernel. After deriving the AWC kernels, we visualized them as sub-filters.

We used Gaussian white noise (mean = 0, std = 1) to estimate the unbiased receptive fields of all convolutional units.

We defined a stimulus as a column vector $\mathbf{s}$ in N-dimensional space,

$$\mathbf{s} \triangleq \begin{pmatrix} s_1 \\ s_2 \\ \vdots \\ s_N \end{pmatrix} \sim N(\mathbf{0}, \mathbf{1}). \tag{1}$$

In this analysis, there were $\mathbf{k}_m$ linear filters,

$$\mathbf{k}_m \triangleq \begin{pmatrix} k_{m,1} \\ k_{m,2} \\ \vdots \\ k_{m,N} \end{pmatrix}, \tag{2}$$

where $m$ is the number of filters ranging from 1 to $M$ depending on the model. Then, $K = span(\mathbf{k}_1, \mathbf{k}_2, \cdots, \mathbf{k}_M)$ is the subspace of the unit as a nonlinear system. Next, we obtained the signals as an inner product of the filter and stimuli, $\mathbf{k}_m^T \mathbf{s}$,

$$\mathbf{k}_m^T \mathbf{s} = \sum_{i=1}^{N} k_{m,i} \cdot s_i. \tag{3}$$

Under the assumption of a LNP model, the signals are nonlinearly transformed into the probability of spiking as a response to the stimulation $\mathbf{s}$. However, the output of our model was the activations of feature maps of VGG16, not spiking responses. Thus, in this study, we did not need a Poisson spiking generator in the response prediction model and accordingly eliminated the Poisson process from the model. Lastly, we estimated the response of a unit, $\mathbf{R}_{unit}$, with a nonlinear function $\varphi$

$$\mathbf{R}_{unit} = \varphi(\mathbf{k}_1^T \mathbf{s}, \mathbf{k}_2^T \mathbf{s}, \cdots, \mathbf{k}_M^T \mathbf{s}). \tag{4}$$

According to equation (3), to identify the nonlinear systems, we needed to know $\mathbf{k}_m$ of all convolutional units in VGG16. We therefore defined $\hat{\mathbf{g}}_k$ as the $k$th approximated Wiener kernel. The responses to Gaussian white noise of VGG16, $\mathbf{R}_{unit}$, was defined as follows:

$$R_{unit} \approx \hat{g}_0 + \hat{g}_1^T s + s^T \hat{g}_2 s. \tag{5}$$

When the stimulus is Gaussian white noise, $\hat{g}_k$ can be expressed as follows: $\hat{g}_0 = \frac{1}{n}\sum_{i=1}^{n} R_i$, $\hat{g}_{1j} = \frac{1}{\sigma^2}\frac{1}{n}\sum_{i=1}^{n} R_i s_{i-j}$, $\hat{g}_{2kl} = \frac{1}{2\sigma^2}\sum_{i=1}^{n} s_{i-k} R_i s_{i-l}$.

Introducing $G_2 \propto s^T s$, because $G_2$ is a symmetric matrix, its eigenvectors $e_i$ form an orthogonal basis. Then, transformation

$$G_2 = \sum_{i=1}^{N} \lambda_i e_i e_i^T \tag{6}$$

is true with eigenvalues $\lambda_i$, and we get

$$R_{unit} \approx \hat{g}_0 + \hat{g}_1^T s + \sum_{i=1}^{N} \lambda_i (s^T e_i)^2. \tag{7}$$

The AWA and AWC matrices are defined as follows:

$$\mathbf{AWA} = \frac{1}{\sum_{i=1}^{n}|R_i|} \sum_{i=1}^{n} R_i s_i. \tag{8}$$

$$\mathbf{AWC} = \frac{1}{\sum_{i=1}^{n}|R_i|} \sum_{i=1}^{n} (R_i s_i - STA)(R_i s_i - STA)^T. \tag{9}$$

In sum, the AWA filter approximates $g_1$, and the product of the STC matrix and the relevant gain approximates $g_2$. Linear filters of the Wiener model and our LN model can be obtained as the AWA filter and eigenvectors of the AWC matrix.

Mathematically, eigenvalue decomposition of the AWC matrix corresponds to a principal component analysis (PCA), that is, the second order Wiener kernel extracts linear filters using the differences of variance. Thus, eigenvectors with larger eigenvalues were considered to be excitatory filters of the unit, and eigenvectors with smaller eigenvalues were thought to be suppressive filters. In this study, this analysis was done with a center $73 \times 73$ pixels field of $224 \times 224$ pixels input stimulus images because of limited computational resources.

### 2.7 A model for evaluating the estimated receptive field structures

After estimating the first and second Wiener kernels, we verified their validity. We implemented a LN model with the Wiener kernels by modifying the Rust model (Rust et al., 2005). There were two types of sub-filters: excitatory and suppressive filters (Fig. 2): the former enhanced the response and the latter suppressed it. We needed to insert these filters into our model as linear filters. However, the number of significant filters may alter among units. To generalize the model, the number of excitatory and suppressive filters was fixed to 10 each, and the weights of each filter were calculated as free parameters by the least squares method. The weighted signals were transformed as follows. For the AWA filter, an identity function was used, and for the AWC sub-filters, a full-wave rectified function was used (Fig. 2). These operations were performed in each RGB channel. All transformed signals were summed as excitatory inputs ($E$) and suppressive inputs ($S$). The response of a unit was defined as follows:

$$\boldsymbol{R_{unit} = \alpha + (\beta E - \gamma S)}, \tag{10}$$

where $\boldsymbol{\alpha}$ is the baseline response, and $\boldsymbol{\beta}$ and $\boldsymbol{\gamma}$ represent the gain of the excitatory and suppressive component, respectively. We fitted this function with normalized responses to Cartesian gratings and natural images. Then the prediction accuracy was defined as a correlation coefficient between actual responses and $\boldsymbol{R_{unit}}$.

We also predicted the measured response with only an AWA filter and the LN model. Because the AWA filter equals the first Wiener kernel, a comparison of the AWA and AWC models corresponds to a comparison between the model including only the first Wiener kernel model and the model including both the first and second Wiener kernels.

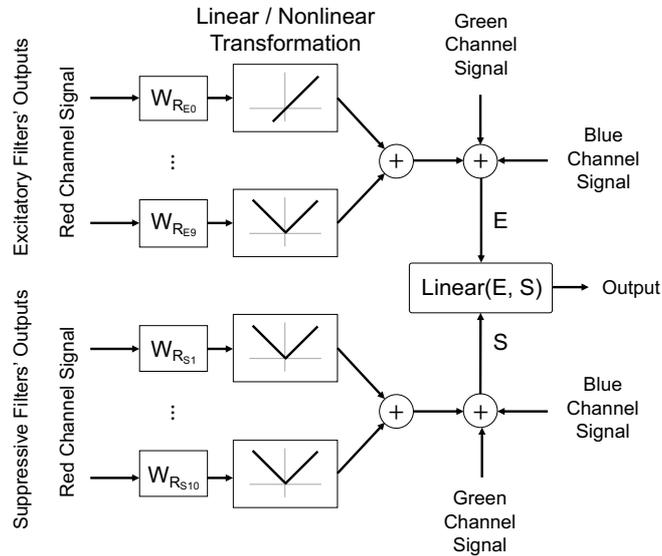

Figure 2. Illustration of the LN model. Twenty (10 excitatory and 10 suppressive) filters were used. One excitatory filter was the AWA filter, and the other filters were obtained from AWC analysis.

# 3 Results

**Table 1. receptive field sizes and unit numbers in each layer.**

| Layer | RF size (pixels) | Number of units |
|---|---|---|
| block1_conv1 | 3 × 3 | 64 |
| block1_conv2 | 5 × 5 | 64 |
| block2_conv1 | 6 × 6 | 128 |
| block2_conv2 | 10 × 10 | 128 |
| block3_conv1 | 12 × 12 | 256 |
| block3_conv2 | 20 × 20 | 256 |
| block3_conv3 | 28 × 28 | 256 |
| block4_conv1 | 32 × 32 | 512 |
| block4_conv2 | 48 × 48 | 512 |
| block4_conv3 | 64 × 64 | 512 |
| block5_conv1 | 64 × 64 | 512 |
| block5_conv2 | 96 × 96 | 512 |
| block5_conv3 | 128 × 128 | 512 |

## 3.1 Reconstructed spatial structure of receptive fields of DCNN units

To reconstruct the receptive field spatial structure of DCNN units, we applied AWA and AWC analyses to VGG16 responses to grating stimuli and natural image stimuli. The sizes of the receptive fields and the number of units are shown in Table 1. The receptive field size of the units increased as the layer progressed, which is consistent with the receptive field size of neurons in the ventral visual pathway of biological neural networks.

Figure 3 shows a reconstructed AWA filter and AWC sub-filters and corresponding eigenvalues at each color channel of unit #2 in block3_conv3. The most prominent spatial structures were observed in the green channel, while those in the blue channel were ambiguous. In general, the spatial structures of the AWA filter and AWC sub-filters were very similar among R, G, and B channels. This trend was true for most of units in VGG16. The spatial structures of the AWC sub-filters were distinctive from that of the AWA filter, because AWA was subtracted when conducting singular value decomposition for the covariance matrix (Eq. 9). Both the AWA filter and AWC sub-filters exhibited spatially segregated on and off regions. The AWA filter tuned to low spatial frequency, and the AWC sub-filters tuned to high spatial frequency. The preferred orientation of all excitatory AWC sub-filters was around 90°(vertical), while the preferred spatial phases of sub-filter #1 and sub-filter #2 were ~180° different. The unit exhibited spatial phase invariance to grating stimuli that corresponded the property of complex cells in V1. In contrast, the preferred orientation of the suppressive filters was ~ 0° (horizontal) and orthogonal to that of the excitatory sub-fields, which means the unit showed cross-orientation inhibition, which is a well-known suppressive property of V1 neurons (Carandini et al., 1998; DeAngelis et al., 1992; Kimura & Ohzawa, 2009). Also noteworthy was that the suppressive sub-filters exhibited an anti-phase preference like the excitatory filters, suggesting the cross-orientation inhibition of this unit was also spatial phase invariant.

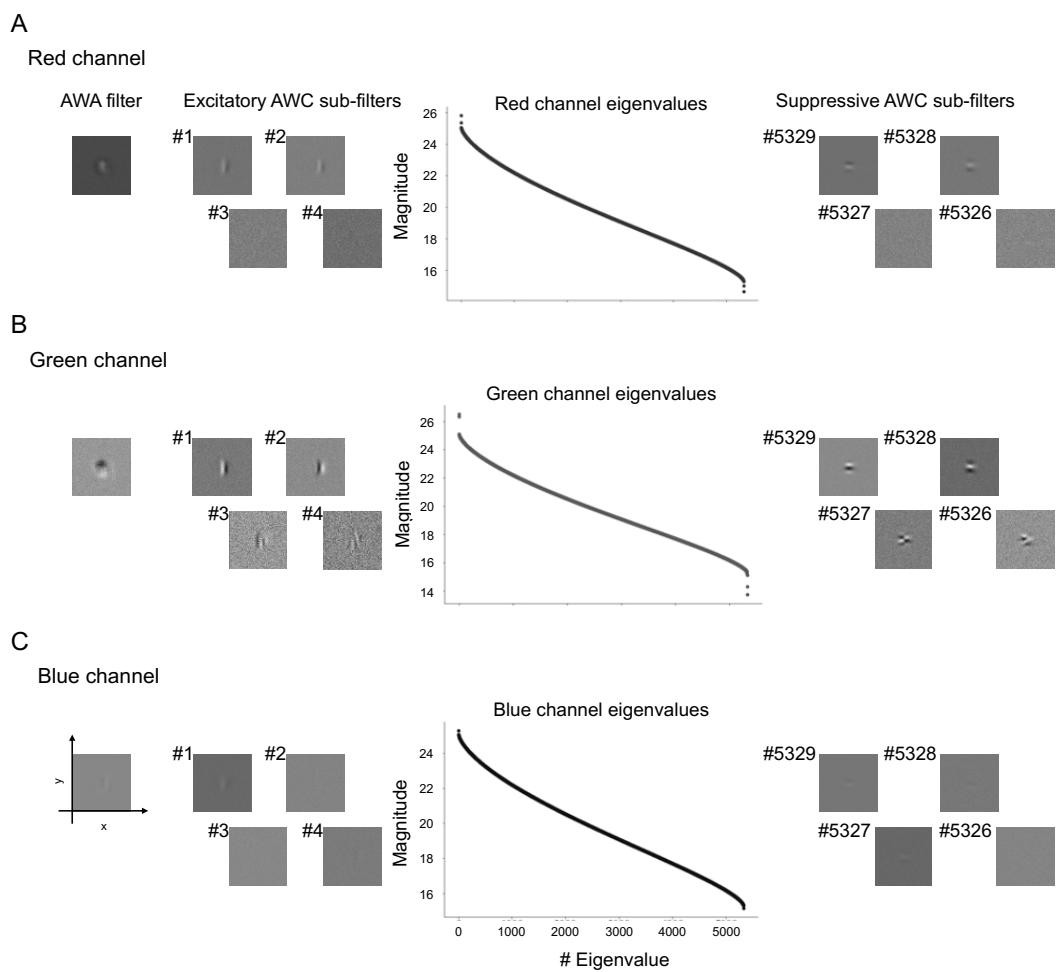

Figure 3. **An example of AWC analysis. The reconstructed AWA filter and AWC sub-filters of a unit in block3_conv3 in the Red (A), Green (B), and Blue (C) channels. The preferred orientations of the excitatory and suppressive sub-filters were orthogonal in all channels.**

Figure 4 shows typical examples of spatial structures of the receptive field of units in block1_conv1 (Fig. 4A), block2_conv1 (Fig. 4B), and block4_conv1 (Fig. 4C). Almost all units in block1_conv1 exhibited a center-surround antagonistic spatial structure or odd Gabor function-like structure, which is consistent with the spatial receptive field of retinal ganglion cells (or relay cells in the lateral geniculate nucleus) and simple cells in V1, respectively. We also observed several units exhibited a uniform color patch-like spatial receptive field (Fig. 4A unit #2). In block1_conv1, almost no units exhibited visible spatial structures of the AWC sub-filters, suggesting only a AWA filter was enough to explain the receptive field of the input layer.

In the middle layer (block2_conv1), the AWA filter exhibited a slightly complicated spatial structure at all color channels (Fig. 4B far left). Nevertheless, the spatial structures of the excitatory AWC sub-filters appeared as a simple Gabor function, but their preferred orientations were slightly different (Fig. 4B second and third columns from left). Therefore, we assumed that this unit exhibits curvature or V-shape selectivity, which is a reported property of V4 neurons in biological networks (R. Tang et al., 2020; Yau et al., 2013). We could not detect a visible spatial structure of suppressive AWC sub-filters in this unit (Fig. 4B first and second column from right).

The results of a block4 unit are shown in Fig. 4C. The unit showed a complicated AWA spatial structure that was hard to interpret (Fig. 4C, far left). In contrast, excitatory AWC sub-filters exhibited a clam shell-like spatial structure with a preference for a small radius of curvature or an acute angle (Fig. 4C, second and third columns from left). The spatial receptive field structure of this unit corresponded well to those of previously reported V4 neurons. The spatial phase of sub-filter #1 and sub-filter #2 differed by ~90°, suggesting the unit showed spatial phase invariance. This unit exhibited suppressive AWC sub-filters, which exhibited a shape that curves or bends inward. The preferred orientation of the suppressive filters was almost orthogonal to that of the excitatory filters, suggesting that the unit also exhibited cross-orientation suppression. The suppressive filters #5329 and #5328 exhibited slightly different spatial phases, suggesting the suppression was spatial phase invariant. In addition, the suppressive filters were finer than the excitatory filters, indicating that they preferred higher spatial frequencies than excitatory filters.

Taken together, as layers progressed, the convolutional units exhibited a more complex spatial receptive field structure, which is consistent with previous reports on neurons in the ventral pathway for visual information processing in monkey and human.

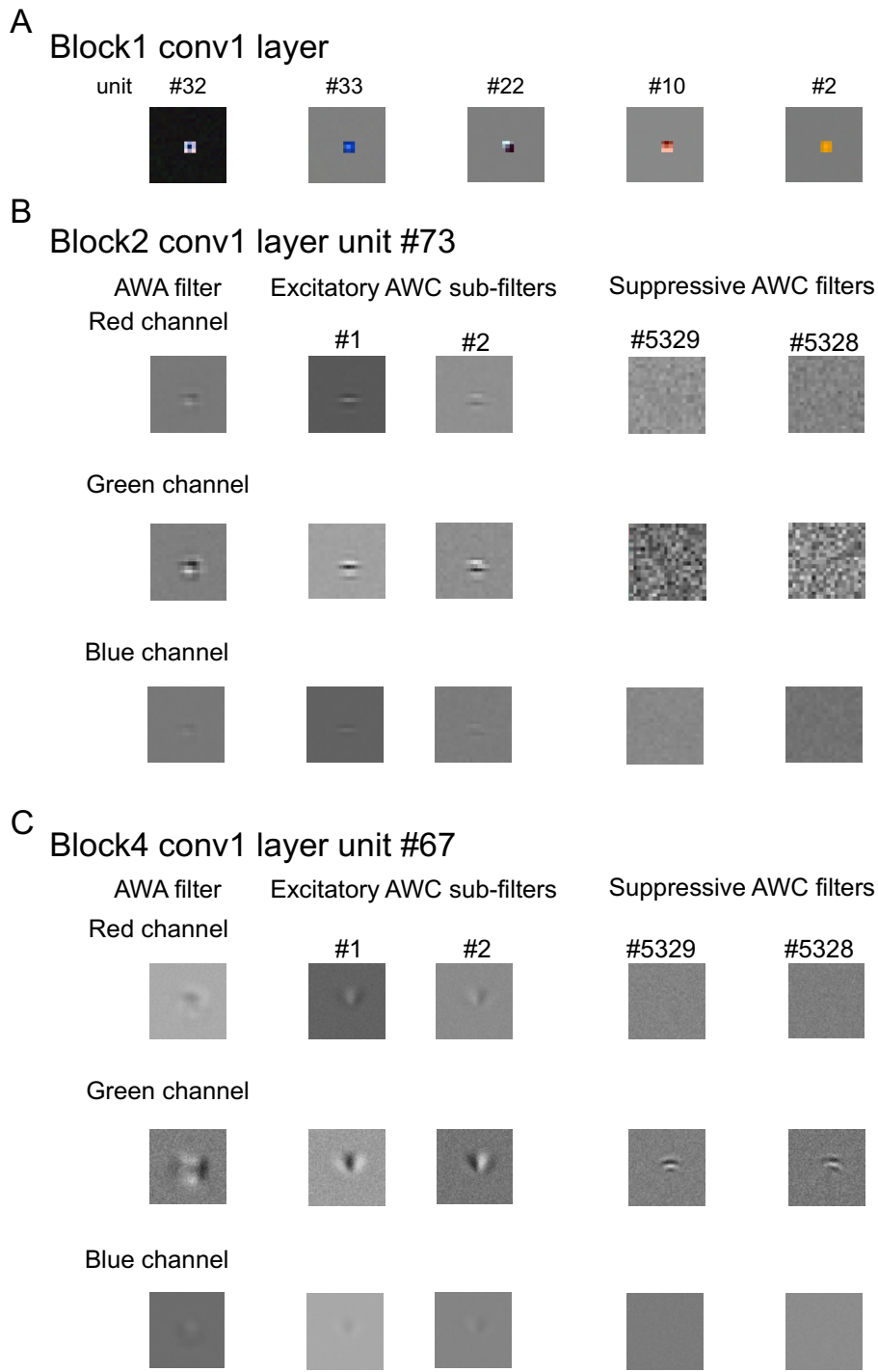

Figure 4. Examples of AWA and AWC analyses. A. AWA filters of units in block1_conv1. B. AWA filters and AWC sub-filters of a unit in block2_conv1. C. AWA filters and AWC sub-filters of a unit in block4_conv1.

### 3.2 Limitations of STC analysis

Figure 5A shows the receptive field structure of a representative unit in block5_conv1. The unit exhibited no visible spatial structures of either the AWA filter or AWC sub-filters. This trend was true for most units in block5_conv1, block5_conv2, and block5_conv3.

Figure 5B shows the responsiveness of the units in 13 convolutional layers to white noise stimuli. Units in block2, block3, and block4 exhibited a heavy tailed distribution, meaning there exists many units that exhibited vigorous responses to white noise stimuli. On the other hand, units in block1_conv1 and all three block5 layers showed a sharp distribution with peaks around zero, meaning white noise did not effectively elicit strong responses by the units in block1 and block5. This result may be attributed to the invisible spatial structures of the AWC sub-filters in block5 (see Discussion section 4.3 Future improvements of AWC analysis).

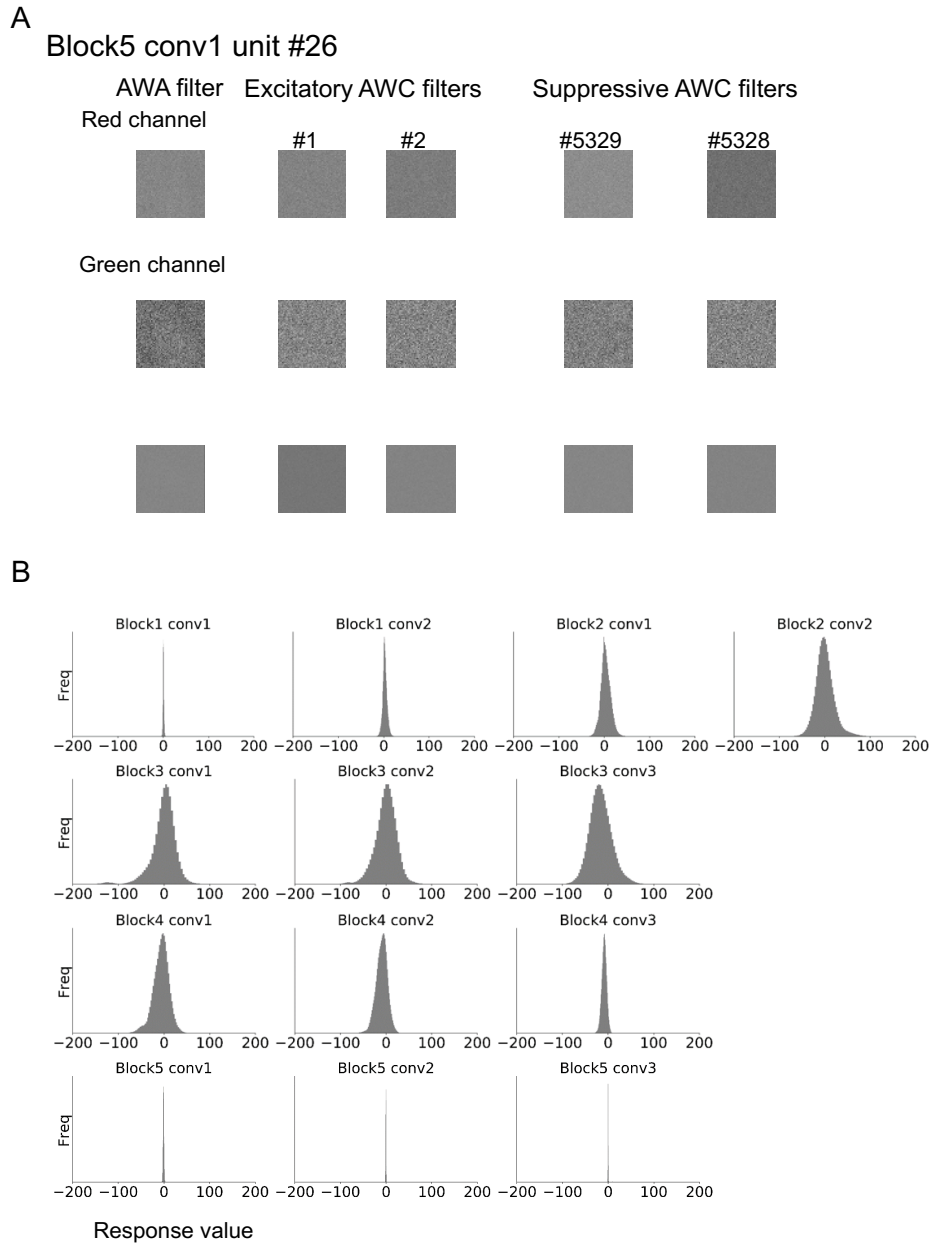

Figure 5. Results of the reverse correlation analysis in block5 and the responsiveness of all layers to white noise stimuli. A. AWA filters and AWC sub-filters of a unit in block5_ conv 1. B. The distribution of responses to white noise stimuli by all units in all layers.

### 3.3 Validity of the AWA filter and AWC sub-filters

To verify the validity of the AWA filter and AWC sub-filters, we calculated the correlation coefficients between the measured unit responses to grating and natural image stimuli and the predicted responses by the AWA filter and AWC sub-filters with the LN model (see 2.7 A model for evaluating the estimated receptive field structures). Figure 6 shows three examples of the prediction accuracy verification. In the input layer (block1_conv1), the unit exhibited monotonic responses to the grating stimuli in the frequency domain (Fig. 6A top row) and no significant image category selectivity (Fig. 6A bottom row). Both the AWA and AWC models exhibited almost perfect response predictions of the convolutional units to grating stimuli and natural image stimuli (Fig. 6A).

In the middle layer (block3_conv1), the unit also exhibited a monotonic response pattern to the grating stimuli (Fig. 6B top row) and no significant image category selectivity (Fig. 6B bottom row). The AWC model still exhibited high prediction accuracy ($r = 0.61$), but the prediction accuracy of the AWA model was far less ($r = 0.15$).

In the higher layer (block5_conv1), the unit exhibited complex orientation and spatial frequency selectivity to the grating stimuli (Fig. 6C top row) and showed significant image category selectivity (Fig. 6C bottom row). This unit exhibited stronger responses to pictures of cars and faces than to bonsais, cups, and roosters (one-way ANOVA, $p < .00000001$). While the prediction accuracy of the AWC model decreased as the layers progressed, the AWC model exhibited a moderate prediction accuracy ($r = 0.39$) that was much higher than that of the AWA model ($r = 0.09$). It is worth mentioning that even though AWC sub-filters were not visible in block5_conv 1 (Fig. 5A), the predicted responses from the AWC model exhibited significant category selectivity ($p < .00000001$) and the same category preference with the measured response (cars and faces).

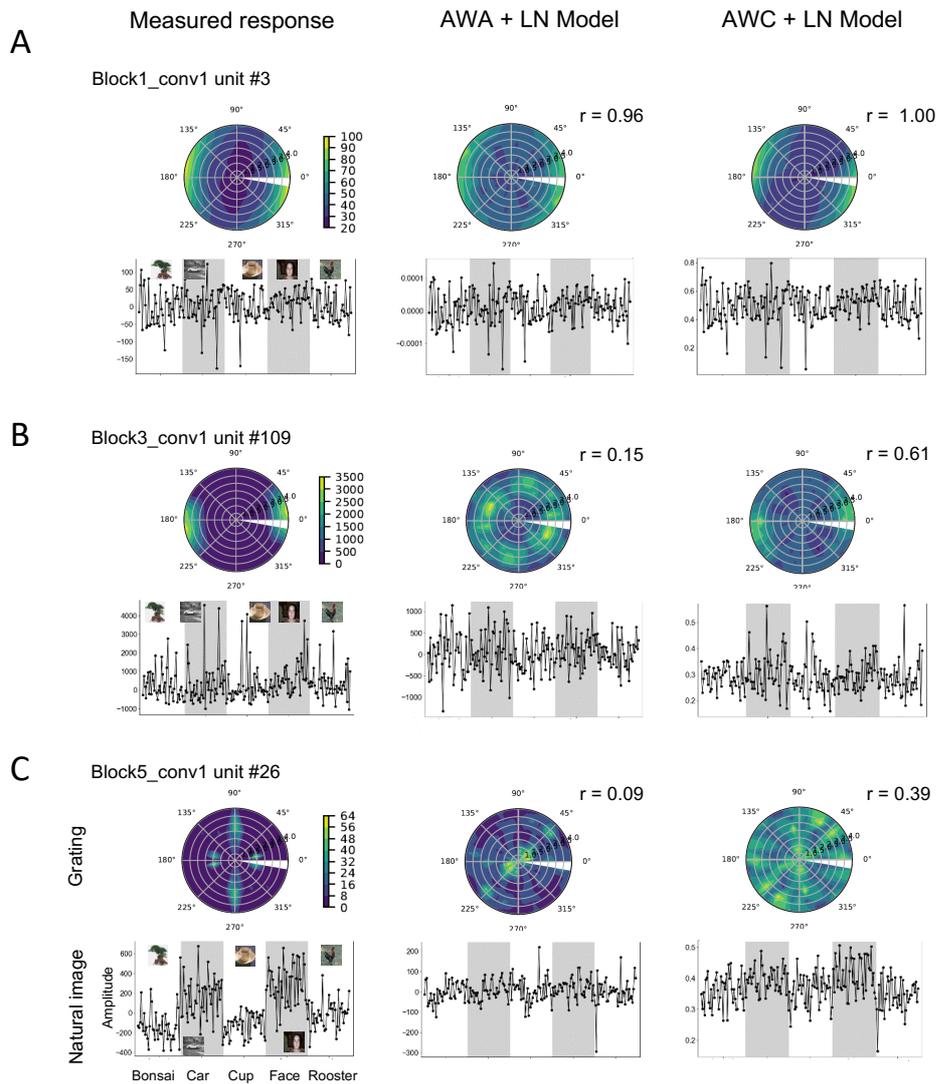

**Figure 6. Prediction accuracy of the AWA and AWC models.** Left: The measured orientation-spatial frequency tuning (top) and responses to natural image stimuli of five different categories (bottom) in block1_ conv1 (A), block3_ conv1 (B), and block5_ conv1 (C). Center and right: the predicted responses from the AWA + LN model (center) and AWC + LN model (right).

Figure 7 shows the mean prediction accuracy of the AWA and AWC models in all convolutional layers. In all layers, the AWC model accuracy was better than that of the AWA model (one-way ANOVA with Bonferroni correction, $p < 0.01$). In block1_conv1 and block1_conv2, the mean prediction accuracies of the AWA and AWC models were much higher (mean of $r > 0.75$ for both models) than the response prediction of V1 neurons with the STC model ($r = 0.56$ for simple cell; $r =$

0.47 for complex cell)(Vintch et al., 2015). We attributed the discrepancy to the internal and measurement noises in the response measurement of V1 neurons, which did not exist in the response measurement of the DCNNs.

The prediction accuracy of both the AWA and AWC models decreased as the layers progressed. In particular, a sharp drop was observed immediately after passing through a max pooling layer (indicated by the half-tone bar in Fig. 7).

In block5, the correlation coefficients of the AWA model were almost 0, meaning that the model did not predict the responses in the higher layer units. In contrast, the AWC model exhibited a significant correlation coefficient $r = \sim 0.2$ even in block5. To verify the significance of the prediction accuracy of the AWC model in block5, we calculated chance level using the same AWC model but replacing the convolutional filters with random noise. Although, the correlation coefficients of the AWC model in block5 were very close to chance level, the difference was significant for block5_conv1 and block5_conv3 ($p < 0.05$).

Taken together, our results suggested that the AWC model explained the fluctuation of the responses of the convolutional units to visual images even in the highest layers, at least to some extent.

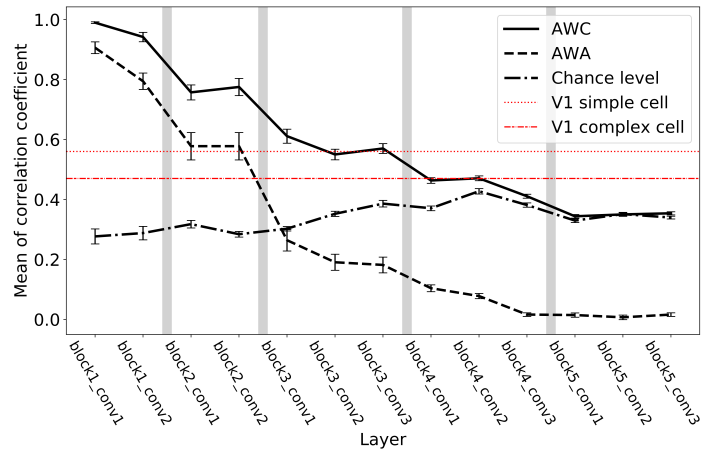

Figure 7. **Prediction accuracy of the AWA and AWC models in all convolutional layers. The vertical and horizontal axes represent mean correlation coefficients and convolutional layers, respectively. Error bars, standard error means of the correlation coefficients. Red dotted and dashed lines, the correlation coefficients of the Rust model for V1 simple and complex cells, respectively (from Vintch et al., 2015). Half-tone bars, location of max pooling layers.**

## 4  Discussion

In this study, we reconstructed the receptive field of convolutional units of VGG16 by reverse correlation techniques (AWA and AWC). The simple LN model with AWC sub-filters captured well the feature selectivity of units in the middle layers (block3 and block4). Even though the visible sub-filters were not reconstructed for block5, the LN model reproduced the natural image category selectivity of the block5 units with the AWC sub-filters.

### 4.1 Validity of non-linear function selection

In this study, we adopted a simple non-linear function in the LN model (Fig. 2), which reconstructed the full-wave rectification of the inner product between the AWC sub-filters and stimuli. Although this simple non-linear function worked well at lower layers (block1 and block2), the prediction accuracy decreased in the middle (block3 and block4) and higher layers (block5). We assumed that the non-linear response was not the same among DCNNs. Indeed, in the input layer (block1_conv1), the AWA model predicted well the unit responses, in which the model acted as a linear model with a linear filter, suggesting that the units in the input layer responded to the stimuli as a linear filter.

However, as the layers progressed, the prediction accuracy of the AWA and AWC models decreased. This result suggests that the nonlinearity becomes stronger as the layers progress. In particular, a sharp drop of the prediction accuracy was observed immediately after passing through a max pooling layer (indicated by the half-tone bar in Fig. 7). In DCNNs, although max pooling is primarily used to reduce the dimensionality of the input to reduce overfitting and use of computational resources, it also is a non-linear operation. The results of our LN model captured the non-linearity caused by the activation function for the outputs from the convolutional layers (ReLU for VGG16), but failed to capture the non-linearity caused by the max pooling layers. Therefore, we should have used distinctive non-linear models before and after a max pooling layer.

### 4.2 Advantages of reverse correlation analysis against activation maximization analysis

The reconstructed AWA filter and AWC sub-filters allowed us to interpret the functional significance of each unit (Fig. 3 and Fig. 4). The unit shown in Fig. 3 was a putative orientation detector that showed position invariance within its receptive field, and the unit shown in Fig. 4B detects curvature or sharp 'V' shapes. The AWC sub-filters of the unit were a Gabor function with slightly different preferred orientations. We assumed this unit receives multiple inputs from units whose receptive fields are V1 simple cell-like Gabor functions with slightly different preferred orientations.

The receptive field of the unit in Fig. 4C was like a cram shell or arch with segregated on-off regions, which can thus exhibit curvature selectivity. The shape of the AWC sub-filters were totally different between block2 and block4. In block4, the receptive field structure of the AWC sub-filters was not a Gabor function but a combination of Gabor filters with different center positions, preferred orientations, and spatial frequencies.

Another important result is that the AWC sub-filters made it possible to analyze excitatory and suppressive receptive field structures independently. The unit in Fig. 4C showed suppressive sub-filters with a shape that curves or bends inward and orthogonal to the excitatory sub-filters. Furthermore, the stripe of the receptive field was thinner than that of excitatory sub-filters, suggesting the suppression was tuned to higher spatial frequencies. Because of the suppression, the unit will show stronger feature selectivity than without suppression, that is, the unit responds well to a 'V' shape but not well to '∀'.

Using receptive field structures, the visual feature selectivity of the units can be analyzed quantitatively. In this way, we can interpret the functional significance of the convolutional units and layers in DCNNs using reconstructed AWC sub-filters. This point gives AWC analysis techniques an important advantage over activation maximization analysis techniques.

### 4.3 Future improvements of AWC analysis

Although AWC analysis successfully reconstructed the receptive field structures of convolutional units in VGG16 from block1 to block4, it failed to visualize the spatial structures of the receptive fields of units in block5. At the same time, the response prediction of the AWC model captured the image category selectivity of the units in block5 (Fig. 6), but that of the AWA model did not. These results suggested that although the spatial structure of the AWC sub-filters were not visible, they contain information about the feature selectivity of the units in block5.

One possible reason why the AWC sub-filters were invisible in block5 is that the elicited responses by white noise stimuli were too weak to effectively reconstruct the AWC sub-filters. As can be seen in Fig. 5B, block1_conv1 and all three layers in block5 showed very weak responses to white noise (most of the responses were concentrated near zero, which indicates no responses). As mentioned above, the receptive field of the unit in block1_conv1 was captured well by a single AWA filter. The AWA analysis requires relatively weak responses compared to the AWC analysis. However, unit #5 in block5 exhibited a significantly complicated feature selectivity for grating stimuli and natural image stimuli (Fig. 6), making the AWC sub-filters necessary to obtain high prediction accuracy with the LN model. Therefore, to visualize reconstructed AWC sub-filters, we should have used stimuli that can elicit vigorous responses of the units in block5 (e.g. whitened natural images). This point is for future studies.

Another possible solution is to use a more elaborated STC analysis (e.g. Bayesian spike-triggered covariance analysis), which could improve the accuracy of the reconstructed receptive field structure with a smaller number of stimulus presentations (Pillow & Park, 2011).

### 4.4 Applications to neuroscience

Recent studies reported similarities between the visual information processing of biological neural networks and DCNNs (Eickenberg et al., 2017; Güçlü & van Gerven, 2015). In the ventral stream of the visual cortex, it was reported that the convolutional units in the top layer of a CNN showed a strong correlation with neurons in the IT cortex (Hong et al., 2016; Khaligh-Razavi & Kriegeskorte, 2014; Yamins et al., 2014). It was also reported that the feature selectivity of the units in the middle layer of the CNN was very similar with that of V4 neurons (Yamins et al., 2014), confirming a decades-old prediction (Lehky & Sejnowski, 1988; Zipser & Andersen, 1988). Other research has shown that DCNNs can reproduce important functions in the visual areas of biological neural networks (Eickenberg et al., 2017; Güçlü & van Gerven, 2015).

Reverse correlation analysis is a powerful tool for understanding the functional significance of a single neuron by visualizing the receptive field structure. However, it is often difficult to collect enough responses from units in the biological neural network for the analysis because of measurement limitations. Assuming DCNNs are a good model for visual areas of biological networks, applying reverse correlation analysis to DCNNs instead will provide important insights on actual neural information processing and information representation in biological neural networks.

Recently, Brain-Score was developed as an index for quantifying the similarity between visual processing in the primate ventral visual pathway and DCNN models (Kubilius et al., 2019; Schrimpf et al., 2020). The Brain-Score benchmark of VGG16 used in this study for V4 is 0.62, which is the highest score among all DCNN models. Accordingly, the receptive field structure of the units shown in Fig. 4C and Fig. 6B give insights on the information representation of V4 neurons in biological neural networks. For insights on the information representation of IT neurons from a reverse correlation analysis of DCNNs, CORnet-S (Kubilus et al.,

2018), whose Brain-Score for IT cortex is 0.423, should be used instead of VGG16, whose Brain-Score for IT cortex is 0.259.

## 5. Conclusion

In this study, we tackled the black box problem of DCNNs by applying reverse correlation analysis. Reconstructed AWC sub-filters with a simple LN model predicted well the unit responses to Cartesian grating stimuli and natural image stimuli in the middle layers (block3 and block4) of VGG16, validating the reconstructed spatial structures of the receptive fields.

The reconstructed receptive fields allowed us to analyze the feature selectivity of the units quantitively, making it possible to clarify the functional significance of the units and layers of DCNNs. This is an advantage of reverse correlation analysis compared with activation maximization analysis.

Our findings also suggested that applying AWC analysis to a DCNN, which provides insights into the information processing and representation by actual neurons in higher visual areas.

**Acknowledgement**
We are grateful to Drs. I. Fujita , I Ohzawa, and P. Karagiannis for their helpful comments on earlier versions of this manuscript and for improving the English. This research and development work was supported by the Ministry of Internal Affairs and Communications Japan and JSPS KAKENHI Grant Number 18H03666, 19K09949 20H04580 and 20K12025 to TN.

**References**

Aljadeff, J., Lansdell, B. J., Fairhall, A. L., & Kleinfeld, D. (2016). Analysis of Neuronal Spike Trains, Deconstructed. *Neuron*, *91*(2), 221–259. https://doi.org/10.1016/j.neuron.2016.05.039

Azevedo, F. A. C., Carvalho, L. R. B., Grinberg, L. T., Farfel, J. M., Ferretti, R. E. L., Leite, R. E. P., Filho, W. J., Lent, R., & Herculano-Houzel, S. (2009). Equal numbers of neuronal and nonneuronal cells make the human brain an isometrically scaled-up primate brain. *Journal of Comparative Neurology*, *513*(5), 532–541. https://doi.org/10.1002/cne.21974

Bonin, V., Mante, V., & Carandini, M. (2005). The Suppressive Field of Neurons in Lateral Geniculate Nucleus. *Journal of Neuroscience*, *25*(47), 10844–10856. https://doi.org/10.1523/JNEUROSCI.3562-05.2005

Carandini, M., Movshon, J. A., & Ferster, D. (1998). Pattern adaptation and cross-orientation interactions in the primary visual cortex. *Neuropharmacology*, *37*(4–5), 501–511. https://doi.org/10.1016/s0028-3908(98)00069-0

Chakraborty, S., Tomsett, R., Raghavendra, R., Harborne, D., Alzantot, M., Cerutti, F., Srivastava, M., Preece, A., Julier, S., Rao, R. M., Kelley, T. D., Braines, D., Sensoy, M., Willis, C. J., & Gurram, P. (2017). Interpretability of deep learning models: A survey of results. *2017 IEEE SmartWorld, Ubiquitous Intelligence Computing, Advanced Trusted Computed, Scalable Computing Communications, Cloud Big Data Computing, Internet of People and Smart City Innovation (SmartWorld/SCALCOM/UIC/ATC/CBDCom/IOP/SCI)*, 1–6. https://doi.org/10.1109/UIC-ATC.2017.8397411

Chichilnisky, E. J. (2001). A simple white noise analysis of neuronal light responses. *Network: Computation in Neural Systems*, *12*(2), 199–213. https://doi.org/10.1080/net.12.2.199.213

Chollet, F., & others. (2015). *Keras*.

DeAngelis, G. C., Ohzawa, I., & Freeman, R. D. (1993). Spatiotemporal organization of simple-cell receptive fields in the cat's striate cortex. II. Linearity of temporal and spatial summation. *Journal of Neurophysiology*, *69*(4), 1118–1135. https://doi.org/10.1152/jn.1993.69.4.1118

DeAngelis, G. C., Robson, J. G., Ohzawa, I., & Freeman, R. D. (1992). Organization of suppression in receptive fields of neurons in cat visual cortex. *Journal of Neurophysiology*, *68*(1), 144–163. https://doi.org/10.1152/jn.1992.68.1.144

Eggermont, J. J., Johannesma, P. I. M., & Aertsen, A. M. H. J. (1983). Reverse-correlation methods in auditory research. *Quarterly Reviews of Biophysics*, *16*(3), 341–414. https://doi.org/DOI: 10.1017/S0033583500005126

Eickenberg, M., Gramfort, A., Varoquaux, G., & Thirion, B. (2017). Seeing it all: Convolutional network layers map the function of the human visual system. *NeuroImage*, *152*(January 2016), 184–194. https://doi.org/10.1016/j.neuroimage.2016.10.001

Erhan, D., Bengio, Y., Courville, A., & Vincent, P. (2009). Visualizing Higher-Layer Features of a Deep Network. *Technical Report, Univeristé de Montréal*.

Fan, F., Xiong, J., & Wang, G. (2020). On Interpretability of Artificial Neural Networks. In *arXiv*.

Fei-Fei, L., Fergus, R., & Perona, P. (2006). One-shot learning of object categories. *IEEE Transactions on Pattern Analysis and Machine Intelligence*, *28*(4), 594–611. https://doi.org/10.1109/TPAMI.2006.79

Ferster, D., & Miller, K. D. (2000). Neural Mechanisms of Orientation Selectivity in the Visual Cortex. *Annual Review of Neuroscience*, *23*(1), 441–471. https://doi.org/10.1146/annurev.neuro.23.1.441

Fukushima, K., & Miyake, S. (1982). Neocognitron: A Self-Organizing Neural Network Model for a Mechanism of Visual Pattern Recognition. In S. Amari & M. A. Arbib (Eds.), *Competition and Cooperation in Neural Nets* (pp. 267–285). Springer Berlin Heidelberg.

Güçlü, U., & van Gerven, M. A. J. (2015). Deep neural networks reveal a gradient in the complexity of neural representations across the ventral stream. *Journal of Neuroscience*, *35*(27), 10005–10014. https://doi.org/10.1523/JNEUROSCI.5023-14.2015

Hartline, H. K. (1940). The receptive fields of optic nerve fibers. *American Journal of Physiology*, *130*, 690–699.

He, K., Zhang, X., Ren, S., & Sun, J. (2016). Deep residual learning for image recognition. *Proceedings of the IEEE Computer Society Conference on Computer Vision and Pattern Recognition*, *2016-Decem*, 770–778. https://doi.org/10.1109/CVPR.2016.90

Hong, H., Yamins, D. L. K., Majaj, N. J., & Dicarlo, J. J. (2016). Explicit information for category-orthogonal object properties increases along the ventral stream. *Nature Neuroscience*, *19*(4), 613–622. https://doi.org/10.1038/nn.4247

Horwitz, G. D., Chichilnisky, E. J., & Albright, T. D. (2005). Blue-Yellow Signals Are Enhanced by Spatiotemporal Luminance Contrast in Macaque V1. *Journal of Neurophysiology*, *93*(4), 2263–2278. https://doi.org/10.1152/jn.00743.2004

Hubel, D. H., & Wiesel, T. N. (1959). Receptive fields of single neurones in the cat's striate cortex. *The Journal of Physiology*, *148*(3), 574–591. https://doi.org/10.1113/jphysiol.1959.sp006308

Hubel, D. H., & Wiesel, T. N. (1968). Receptive fields and functional architecture of monkey striate cortex. *The Journal of Physiology*, *195*(1), 215–243. https://doi.org/10.1113/jphysiol.1968.sp008455

Jones, J. P., & Palmer, L. A. (1987). The two-dimensional spatial structure of simple receptive fields in cat striate cortex. *Journal of Neurophysiology*, *58*(6), 1187–1211. https://doi.org/10.1152/jn.1987.58.6.1187

Khaligh-Razavi, S. M., & Kriegeskorte, N. (2014). Deep Supervised, but Not Unsupervised, Models May Explain IT Cortical Representation. *PLoS Computational Biology*, *10*(11). https://doi.org/10.1371/journal.pcbi.1003915

Kimura, R., & Ohzawa, I. (2009). Time course of cross-orientation suppression in the early visual cortex. *Journal of Neurophysiology*, *101*(3), 1463–1479. https://doi.org/10.1152/jn.90681.2008

Krizhevsky, A., Sutskever, I., & Hinton, G. E. (2012). ImageNet Classification with Deep Convolutional Neural Networks. In F. Pereira, C. J. C. Burges, L. Bottou, & K. Q. Weinberger (Eds.), *Advances in Neural Information Processing Systems 25* (pp. 1097–1105). Curran Associates, Inc. http://papers.nips.cc/paper/4824-imagenet-classification-with-deep-convolutional-neural-networks.pdf

Kubilius, J., Schrimpf, M., Kar, K., Rajalingham, R., Hong, H., Majaj, N. J., Issa, E. B., Bashivan, P., Prescott-Roy, J., Schmidt, K., Nayebi, A., Bear, D., Yamins, D. L. K., & DiCarlo, J. J. (2019). Brain-like object recognition with high-performing shallow recurrent ANNs. *Advances in Neural Information Processing Systems*, *32*(NeurIPS), 1–12.

LeCun, Y., Boser, B., Denker, J. S., Henderson, D., Howard, R. E., Hubbard, W., & Jackel, L. D. (1989). Backpropagation Applied to Handwritten Zip Code Recognition. *Neural Comput.*, *1*(4), 541–551. https://doi.org/10.1162/neco.1989.1.4.541

Lehky, S. R., & Sejnowski, T. J. (1988). Network model of shape-from-shading: Neural function arises from both receptive and projective fields. *Nature*, *333*(6172), 452–454. https://doi.org/10.1038/333452a0

Meister, M., Pine, J., & Baylor, D. A. (1994). Multi-neuronal signals from the retina: acquisition and analysis. *Journal of Neuroscience Methods*, *51*(1), 95–106. https://doi.org/https://doi.org/10.1016/0165-0270(94)90030-2

Nguyen, A, Clune, J., Bengio, Y., Dosovitskiy, A., & Yosinski, J. (2017). Plug Play Generative Networks: Conditional Iterative Generation of Images in Latent Space. *2017 IEEE Conference on Computer Vision and Pattern Recognition (CVPR)*, 3510–3520. https://doi.org/10.1109/CVPR.2017.374

Nguyen, Anh, Dosovitskiy, A., Yosinski, J., Brox, T., & Clune, J. (2016). Synthesizing the Preferred Inputs for Neurons in Neural Networks via Deep Generator Networks. *Proceedings of the 30th International Conference on Neural Information Processing Systems*, 3395–3403.

Pillow, J. W., & Park, I. (2011). Bayesian spike-triggered covariance analysis. *Advances in Neural Information Processing Systems*, 1692–1700. http://machinelearning.wustl.edu/mlpapers/paper_files/NIPS2011_0954.pdf

Pillow, J. W., & Simoncelli, E. P. (2006). Dimensionality reduction in neural models: An information-theoretic generalization of spike-triggered average and covariance analysis. *Journal of Vision*, *6*(4), 9. https://doi.org/10.1167/6.4.9

Reid, R. C., & Alonso, J. M. (1995). Specificity of monosynaptic connections from thalamus to visual cortex. *Nature*, *378*(6554), 281–284. https://doi.org/10.1038/378281a0

Richards, B. A., Lillicrap, T. P., Beaudoin, P., Bengio, Y., Bogacz, R., Christensen, A., Clopath, C., Costa, R. P., de Berker, A., Ganguli, S., Gillon, C. J., Hafner, D., Kepecs, A., Kriegeskorte, N., Latham, P., Lindsay, G. W., Miller, K. D., Naud, R., Pack, C. C., … Kording, K. P. (2019). A deep learning framework for neuroscience. *Nature Neuroscience*, *22*(11), 1761–1770. https://doi.org/10.1038/s41593-019-0520-2

Russakovsky, O., Deng, J., Su, H., Krause, J., Satheesh, S., Ma, S., Huang, Z., Karpathy, A., Khosla, A., Bernstein, M., Berg, A. C., & Fei-Fei, L. (2015). ImageNet Large Scale Visual Recognition Challenge. *International Journal of Computer Vision (IJCV)*, *115*(3), 211–252. https://doi.org/10.1007/s11263-015-0816-y

Rust, N. C., Schwartz, O., Movshon, J. A., & Simoncelli, E. P. (2005). Spatiotemporal elements of macaque V1 receptive fields. *Neuron*, *46*(6), 945–956. https://doi.org/10.1016/j.neuron.2005.05.021

Sakai, H. M. (1992). White-noise analysis in neurophysiology. *Physiological Reviews*, *72*(2), 491–505. https://doi.org/10.1152/physrev.1992.72.2.491

Sakai, H. M., & Naka, K. (1987). Signal transmission in the catfish retina. V. Sensitivity and circuit. *Journal of Neurophysiology*, *58*(6), 1329–1350. https://doi.org/10.1152/jn.1987.58.6.1329

Schrimpf, M., Kubilius, J., Lee, M. J., Ratan Murty, N. A., Ajemian, R., & DiCarlo, J. J. (2020). Integrative Benchmarking to Advance Neurally Mechanistic Models of Human Intelligence. *Neuron*, *108*(3), 413–423. https://doi.org/10.1016/j.neuron.2020.07.040

Schwartz, O, Chichilnisky, E. J., & Simoncelli, E. P. (2002). Characterizing neural gain control using spike-triggered covariance. *Advances in Neural Information Processing Systems 14, Vols 1 and 2*, *14*, 269–276. https://doi.org/10.1016/0041-0101(88)90297-8

Schwartz, Odelia, Pillow, J. W., Rust, N. C., & Simoncelli, E. P. (2006). Spike-triggered neural characterization. *Journal of Vision*, *6*(4), 13. https://doi.org/10.1167/6.4.13

Simoncelli, E., Pillow, J. W., Paninski, L., & Schwartz, O. (2004). Characterization of neural responses with stochastic stimuli. In M. Gazzaniga (Ed.), *The cognitive neurosciences, III* (pp. 327–338). MIT Press.

Simonyan, K., & Zisserman, A. (2014). Very Deep Convolutional Networks for Large-Scale Image Recognition. *ArXiv E-Prints*, arXiv:1409.1556.

Suematsu, N., Naito, T., Miyoshi, T., Sawai, H., & Sato, H. (2013). Spatiotemporal receptive field structures in retinogeniculate connections of cat. *Frontiers in Systems Neuroscience*, *7*, 103. https://doi.org/10.3389/fnsys.2013.00103

Suematsu, N., Naito, T., & Sato, H. (2012). Relationship between orientation sensitivity and spatiotemporal receptive field structures of neurons in the cat lateral geniculate nucleus. *Neural Networks*, *35*, 10–20. https://doi.org/10.1016/j.neunet.2012.06.008

Szegedy, C., Wei Liu, Yangqing Jia, Sermanet, P., Reed, S., Anguelov, D., Erhan, D., Vanhoucke, V., & Rabinovich, A. (2015). Going deeper with convolutions. *2015 IEEE Conference on Computer Vision and Pattern Recognition (CVPR)*, 1–9. https://doi.org/10.1109/CVPR.2015.7298594

Tanaka, K. (1996). Inferotemporal Cortex and Object Vision. *Annual Review of Neuroscience*, *19*(1), 109–139. https://doi.org/10.1146/annurev.ne.19.030196.000545

Tang, R., Song, Q., Li, Y., Zhang, R., Cai, X., & Lu, H. D. (2020). Curvature-processing domains in primate V4. *ELife*, *9*, e57502. https://doi.org/10.7554/eLife.57502

Tang, Y., Nyengaard, J. R., De Groot, D. M. G., & Gundersen, H. J. G. (2001). Total regional and global number of synapses in the human brain neocortex. *Synapse*, *41*(3), 258–273. https://doi.org/10.1002/syn.1083

Touryan, J., Lau, B., & Dan, Y. (2002). Isolation of Relevant Visual Features from Random Stimuli for Cortical Complex Cells. *The Journal of Neuroscience*, *22*(24), 10811 LP – 10818. https://doi.org/10.1523/JNEUROSCI.22-24-10811.2002

Vintch, B., Movshon, J. A., & Simoncelli, E. P. (2015). A Convolutional Subunit Model for Neuronal Responses in Macaque V1. *Journal of Neuroscience*, *35*(44), 14829–14841. https://doi.org/10.1523/JNEUROSCI.2815-13.2015

Yamins, D. L. K., Hong, H., Cadieu, C. F., Solomon, E. A., Seibert, D., & DiCarlo, J. J. (2014). Performance-optimized hierarchical models predict neural responses in higher visual cortex. *Proceedings of the National Academy of Sciences*, *111*(23), 8619–8624. https://doi.org/10.1073/pnas.1403112111

Yau, J. M., Pasupathy, A., Brincat, S. L., & Connor, C. E. (2013). Curvature processing dynamics in macaque area V4. *Cerebral Cortex (New York, N.Y. : 1991)*, *23*(1), 198–209. https://doi.org/10.1093/cercor/bhs004

Zhang, Q., & Zhu, S. (2017). *Visual Interpretability for Deep Learning: a Survey*.

Zhou, B., Khosla, A., Lapedriza, A., Oliva, A., & Torralba, A. (2015). *Object Detectors*

*Emerge in Deep Scene CNNs*.

Zipser, D., & Andersen, R. A. (1988). A back-propagation programmed network that simulates response properties of a subset of posterior parietal neurons. *Nature*, *331*(6158), 679–684. https://doi.org/10.1038/331679a0